\documentclass[conference]{IEEEtran}

\ifCLASSINFOpdf
\else
\fi

\usepackage{amsmath}       
\usepackage{graphicx}      
\usepackage{amsfonts}      
\usepackage{amssymb}       
\usepackage{booktabs}      
\usepackage{array}         
\usepackage{multirow}      
\usepackage{multicol}      
\usepackage{geometry}      
\usepackage{hyperref}      
\hypersetup{colorlinks=true, linkcolor=blue, urlcolor=blue}  
\usepackage{tabularx}      

\begin{document}

\title{Advanced Flood Prediction with Physics-Guided Deep Learning: Combining UNet, FNO, and SAR/Optical Imagery}

\author{
\IEEEauthorblockN{
Tewodros Syum Gebre\IEEEauthorrefmark{1},
Jagrati Talreja\IEEEauthorrefmark{1},
Leila Hashemi-Beni\IEEEauthorrefmark{1}\IEEEauthorrefmark{2}\thanks{Corresponding author: lhashemibeni@ncat.edu}
}
\IEEEauthorblockA{
\IEEEauthorrefmark{1}Built Environment Department, College of Science and Technology,\\
North Carolina A\&T State University, Greensboro, NC 27405, USA\\
Emails: tsgebre@ncat.edu, jtalreja@ncat.edu, lhashemibeni@ncat.edu
}
\IEEEauthorblockA{
\IEEEauthorrefmark{2}United Nations University Institute for Water, Environment and Health,\\
Richmond Hill, ON, Canada
}
}

\maketitle

\begin{abstract}
Accurate and scalable flood mapping remains challenging due to limited ground observations, heterogeneous terrain conditions, and the difficulty of enforcing hydrodynamic consistency within data-driven models. This work introduces a physics-guided deep learning framework that integrates multi-modal remote sensing (Sentinel-1 SAR, Sentinel-2 optical imagery, and DEM-derived terrain features) with constraints from the depth-averaged shallow water equations (SWE). The proposed hybrid architecture combines a UNet to capture fine-scale spatial details with a Fourier Neural Operator (FNO) to model basin-scale hydraulic interactions, while physics-informed residual losses ensure mass and momentum consistency. Evaluated across diverse floodplain settings, the hybrid model achieves an Intersection over Union of 0.82 and an F1 score of 0.90 for flood extent prediction, outperforming UNet-only and FNO-only baselines. Using hydrodynamic simulations as reference data, the model achieves an RMSE of 0.21 m for water depth and 0.15 m/s for flow velocity. Physics consistency is maintained, with low residuals and mass imbalance below 2.1\%. Ablation studies confirm that removing physics-based regularization significantly degrades performance, underscoring the value of physical constraints for stability and generalization. These results demonstrate that embedding hydrodynamic principles into deep learning yields more accurate, reliable, and physically coherent flood predictions, offering strong potential for operational monitoring and large-scale deployment. 

This paper has been accepted for publication in the Proceedings of the IEEE Radar Conference (RadarConf 2026). The final authenticated version will be available through IEEE Xplore.
\end{abstract}

\IEEEpeerreviewmaketitle

\section{Introduction}

Floods are among the most damaging natural hazards worldwide. Their frequency and severity are rising due to climate change, rapid urbanization, and intense rainfall. Recent advances in satellite remote sensing, especially frequent SAR and multispectral optical observations, offer new opportunities for rapid flood mapping \cite{peng2025automatic}. Yet, turning these observations into accurate and physically meaningful flood information remains challenging \cite{zhang2025sar}.

The core issue is the gap between what satellites measure and what decision-makers need. Sensors capture raw observations (e.g., digital numbers) which are processed into backscatter, reflectance, and coherence, while emergency response depends on water depth, surface elevation, and flow velocity. Estimating these variables from imagery is difficult because of sensor noise, cloud cover, temporal gaps, and the nonlinear nature of hydrodynamic processes. The problem is worse in data-scarce regions where field measurements and high-resolution hydraulic models are limited.

This work proposes a physics-guided deep learning framework that combines multi-modal satellite data with the shallow water equations. Unlike conventional flood mapping methods that rely only on classification or purely data-driven regression, our approach embeds physical laws into the learning process. We integrate a U-Net (UNet) for fine-scale spatial features with a Fourier Neural Operator (FNO) for basin-scale hydrodynamics. This hybrid design predicts continuous hydrodynamic state variables while maintaining physical consistency. Our main contributions include: (1) a hybrid UNet–FNO architecture that captures both local inundation patterns and large-scale flow dynamics; and (2) a physics-guided training approach leveraging shallow water equation residuals to ensure consistency with minimal supervision. Unlike pure emulators that rely solely on simulation data, our framework actively assimilates SAR-derived flood extent. The physics loss acts as a regularization term, constraining the solution space to physically plausible states, while the SAR input forces the model to respect observed boundaries that static hydraulic models might miss due to outdated bathymetry.

The rest of the paper is organized as follows: Section II reviews related work on remote sensing, deep learning, and physics-informed methods. Section III details the proposed architecture and training approach. Section IV describes the experimental results and discusses implications and limitations. Section V concludes and outlines future directions.

\section{Related Works}

Flood mapping using remote sensing has evolved from threshold-based classification to advanced learning-based approaches. SAR imagery has been widely adopted due to its all-weather capability. Classical methods such as Otsu thresholding, region-growing, and Bayesian frameworks have been extensively applied for flood delineation in urban areas \cite{giustarini2012change}. Despite these advances, SAR-based mapping faces challenges including speckle noise, surface roughness ambiguity, and layover/shadow effects \cite{amitrano2024flood}. Recent studies have introduced machine learning and fuzzy logic to improve SAR flood detection \cite{tanoh2023spatial,fawakherji2025flood,jamali2024residual,pillai2024flood,fawakherji2025deepflood}. Our work extends these efforts by predicting continuous hydrodynamic fields rather than binary flood masks, leveraging SAR as a primary modality.

Optical imagery has traditionally relied on water indices such as NDWI and MNDWI \cite{bormudoi2024common}. However, cloud cover and temporal gaps limit its reliability during flood events. Deep learning models, particularly CNNs and U-Net variants, have improved segmentation accuracy using Sentinel-2 imagery \cite{konapala2021exploring,fawakherji2023multichannel,blay2024flood}. Our approach fuses optical imagery with SAR to exploit complementary strengths, ensuring robustness under cloud-obscured conditions.

Digital Elevation Models (DEMs) provide terrain-derived indicators such as HAND, slope, and curvature, which are widely used for flood susceptibility mapping \cite{albano2025use}. DEM conditioning enhances flow direction and supports hydrogeomorphic flood prediction. In our framework, DEM features inform SWE-based source terms (bed slope, friction) in the physics-guided formulation.

U-Net and its variants dominate flood segmentation tasks, often incorporating attention mechanisms and multi-scale encoders. Multi-sensor fusion (SAR + optical) has shown promise \cite{pech2023flooded,fawakherji2024multi}, but most models output binary flood masks without physical interpretability. Existing architectures rarely predict physically consistent hydraulic states.

Efforts to regress water depth or velocity from imagery typically rely on supervised learning with synthetic hydraulic simulations \cite{zhang2024estimating}. These approaches often lack physical constraints, leading to unrealistic outputs. Our framework predicts depth and velocity fields under SWE residual penalties, ensuring physical plausibility.

Physics-Informed Neural Networks (PINNs) have been applied to incorporate physical laws including SWE and Navier–Stokes systems into machine learning models \cite{gebre2024ai,gebre2024integrated,gebre2025smart,gebre2025real,tian2025physics}. While PINNs enforce PDE residuals, they suffer from slow convergence and instability near wet–dry fronts. Recent works combine operator learning (e.g., FNO) with physics constraints for scalable PDE surrogates \cite{el2025machine}. Our approach adopts a physics-guided, not fully physics-enforced, strategy using residual penalties and mesh-free sampling, enabling scalability to high-resolution imagery.

FNO has emerged as a powerful tool for learning PDE operators in climate, ocean, and flood modeling \cite{sun2023rapid}. It offers global receptive fields but struggles with fine-grained spatial details. Hybrid CNN–FNO designs leverage local feature extraction and global operator learning \cite{alesiani2022hyperfno}. Our UNet–FNO hybrid captures local flood boundaries while modeling basin-scale hydrodynamics, ensuring consistency with SWE constraints.

The shallow water equations underpin most flood forecasting systems, assuming hydrostatic pressure and depth-averaged velocity \cite{delis2021shallow}. Neural emulators of SWE accelerate simulations but typically require large synthetic datasets and do not assimilate real observations \cite{bentivoglio2023rapid}. Our model directly predicts SWE-consistent states from SAR, optical, and DEM inputs without relying on precomputed simulation libraries.

Existing literature reveals critical gaps: lack of physics integration in remote sensing-based flood mapping, limited ability to infer continuous hydraulic fields, difficulty in multi-modal sensor fusion at scale, PINNs’ inefficiency for large grids, and absence of hybrid architectures combining local and global reasoning. Our contributions include a UNet–FNO hybrid for multi-scale hydrodynamic reasoning, physics-guided learning via SWE residuals, multi-modal ingestion of SAR, optical, and DEM features, scalable continuous flood state estimation, and mesh-free residual sampling aligned with image grids.

\section{Methodology}
\label{sec:method}

We propose a physics-guided deep learning framework for flood mapping and hydraulic state estimation using SAR, optical imagery, and DEM-derived terrain features. The approach integrates remote sensing observations with the depth-averaged shallow water equations (SWE) to enforce physical consistency while leveraging the representational power of neural networks. The key assumptions, steady-state flow, hydrostatic pressure, incompressibility, and depth-averaged velocity, simplify the governing dynamics and enable efficient training under limited observations. Physics constraints are incorporated through residual-based penalties, while data fidelity is enforced via supervised losses on flood extent and optional velocity fields. The overall pipeline combines image-based feature extraction with PDE-informed operators for spatial reasoning.

\subsection{Model Architecture}
\label{sec:model-architecture}

\begin{figure}
    \centering
    \includegraphics[width=1\linewidth]{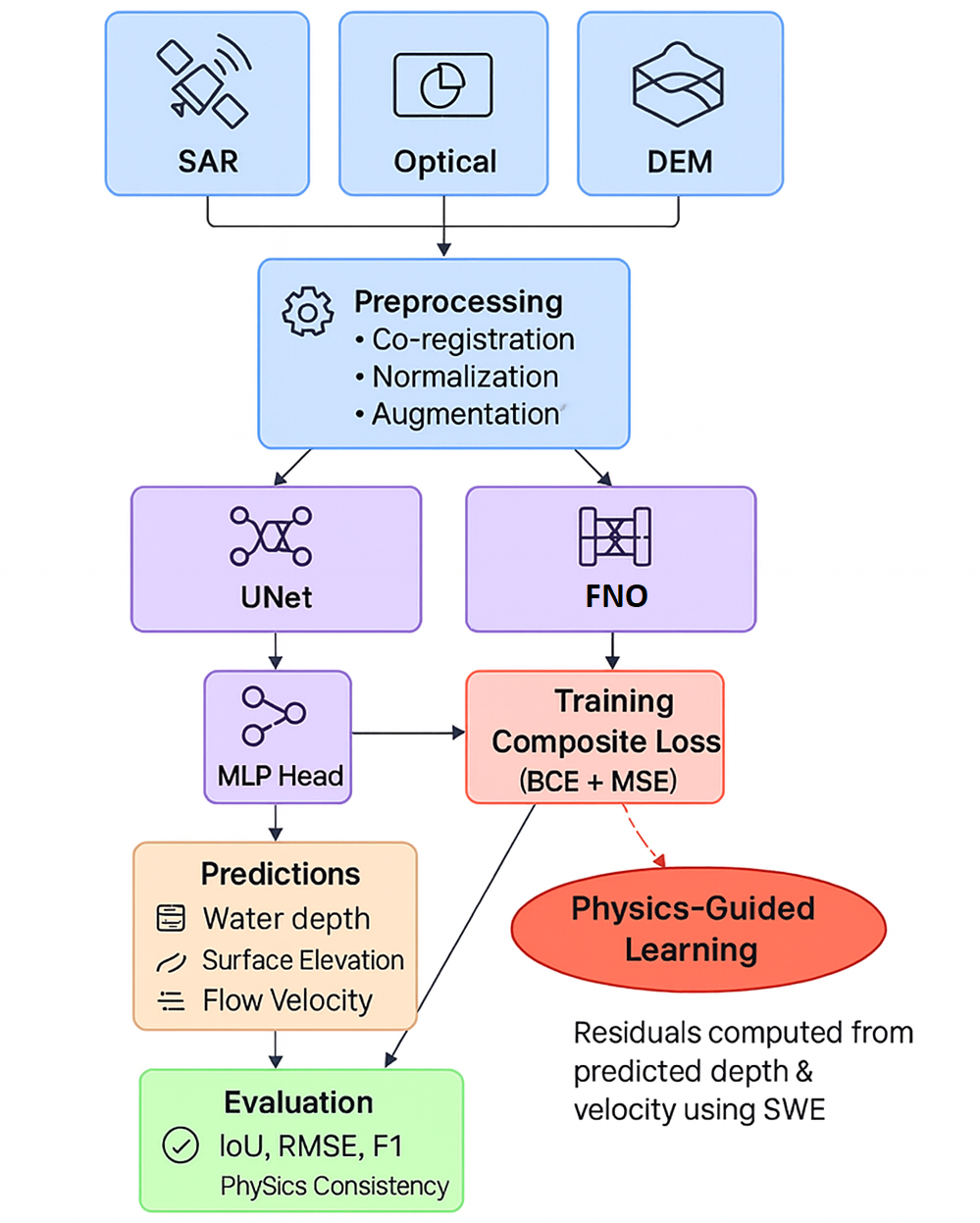}
    \caption{Hybrid architecture combining UNet for local feature extraction and FNO for global hydrodynamic patterns, fused to predict continuous flood variables.}
    \label{fig:pipeline}
\end{figure}

Our architecture is designed to integrate multi-modal remote sensing inputs with physics-informed constraints, enabling accurate and physically consistent flood predictions. It combines two complementary components, the UNet and the Fourier Neural Operator (FNO), to capture local and global patterns in flood dynamics (Figure \ref{fig:pipeline}). This hybrid design ensures that fine-grained spatial details from SAR and optical imagery are preserved while large-scale hydrodynamic interactions are effectively modeled.

\paragraph{Multi-Modal Inputs.}
The model ingests co-registered Sentinel-1 SAR backscatter, Sentinel-2 optical bands (with cloud and shadow masks), and terrain derivatives from DEM (elevation, slope, HAND). These inputs provide complementary information:
\begin{itemize}
    \item \textbf{SAR}: Sensitive to water surfaces and roughness, which enables detection of flood extent under clouds.
    \item \textbf{Optical}: Adds spectral cues for land cover and water presence in clear-sky conditions.
    \item \textbf{DEM}: Provides topographic context for flow paths and hydrostatic pressure terms.
\end{itemize}
This multi-modal fusion is critical for robust flood mapping across diverse landscapes and observation conditions.

\paragraph{Why UNet + FNO?}
Flood dynamics exhibits both local and global dependencies:
\begin{itemize}
\item \textbf{Local Scale}: Accurate flood boundaries, urban features, and small-scale terrain variations necessitate high-resolution spatial detail to capture fine-grained dynamics.
\item \textbf{Global Scale}: Basin-wide water redistribution and upstream-downstream interactions demand long-range modeling to account for broader hydrological processes and connectivity.
\end{itemize}
UNet excels at local feature extraction through its encoder–decoder structure and skip connections, preserving fine spatial details. FNO complements this by operating in the frequency domain, learning low-frequency components that represent large-scale hydrodynamic patterns. This synergy allows the model to capture multi-scale processes that neither component alone could fully represent.

\paragraph{Feature Fusion Strategy.}
Outputs from UNet and FNO are fused using a multi-resolution strategy that combines residual connections and feature concatenation. This preserves UNet’s fine-grained spatial features while integrating FNO’s global context. The fused representation supports downstream prediction of continuous hydrodynamic variables.

\paragraph{Final Output Layer.}
The fused features are passed through a Multi-Layer Perceptron (MLP) head to predict water depth, surface elevation, and depth-averaged velocity $(u,v)$ at each spatial location. The choice of an MLP is motivated by the need for continuous outputs rather than discrete classes. This design also facilitates integration with physics-informed constraints, as scalar outputs can be directly used in shallow water equation residuals.

\paragraph{Connection to Physics-Guided Learning.}
Unlike purely data-driven models, our architecture is explicitly designed to support physics-based regularization. The continuous outputs from the MLP enable computation of residuals from the depth-averaged SWE (Section \ref{sec:phys-train-eval}). This coupling ensures that predictions are not only accurate with respect to observed flood extent but also consistent with hydrodynamic principles, improving generalization under sparse or noisy observations.

\subsection{Physics-Guided Learning with Remote Sensing Constraints}
\label{sec:phys-train-eval}

Our approach leverages the depth-averaged shallow water equations (SWE) as a physics backbone, combined with SAR, optical imagery, and DEM-derived topography to guide learning. These physical assumptions simplify the governing dynamics while preserving essential hydrologic behavior, enabling efficient training under limited observations. Table \ref{tab:assumptions} summarizes the assumptions, their implementation, and computational impact.

\begin{table}[ht]
\small
\centering
\caption{Physical assumptions for efficient physics-constrained learning.}
\label{tab:assumptions}
\begin{tabularx}{\columnwidth}{>{\hsize=1.2\hsize}X >{\hsize=0.9\hsize}X >{\hsize=1\hsize}X}
\hline
\textbf{Assumption} & \textbf{Implementation} & \textbf{Impact} \\
\hline
Steady-state flow & Remove time derivatives from SWE & Simplifies residuals; enables spatial-only autodiff \\
Hydrostatic pressure & Drop vertical momentum equations & 2D depth-averaged formulation \\
Depth-averaged velocity & Predict $(u,v)$ per point & Matches SAR-derived velocities; reduces parameters \\
Incompressibility & Treat $\rho$ as constant & Simplifies momentum equations \\
No-slip at bed & Use terrain slope ($\nabla z_b$) in source terms & Encodes bed friction implicitly \\
Mesh-free method & Sample residuals continuously & Adaptive, grid-free training; improves generalization \\
\hline
\end{tabularx}
\end{table}

\paragraph{Data–Physics Coupling.}
Available observations include:
\begin{itemize}
    \item \textbf{Flood extent} from SAR/optical classification (binary mask).
    \item \textbf{Surface velocity proxies} from SAR coherence or offset tracking (optional).
    \item \textbf{Topography} derived from Copernicus GLO-30 DEM for bed elevation $z_b$ and slope $\nabla z_b$.
\end{itemize}
Flood extent defines the wet mask $\Omega_w = \{(x,y): h > \varepsilon\}$ and drives $\mathcal{L}_{\text{ext}}$. DEM-derived slopes support hydrostatic and friction terms, while SAR-derived velocities constrain $(u,v)$ predictions when available.

\paragraph{Governing Equations.}
Let $h(x,y)$ denote water depth, $\mathbf{u}=(u,v)$ the depth-averaged horizontal velocity, $z_b(x,y)$ the bed elevation, and $\eta=z_b+h$ the free-surface elevation. Gravity is $g$. Assuming hydrostatic pressure and incompressibility, the depth-averaged SWE simplify to:
\begin{align}
    \partial_t h + \nabla \cdot (h\mathbf{u}) &= 0, \tag{1}\label{eq:mass} \\
    \partial_t (h u) + \nabla \cdot (h u \mathbf{u}) + g h \,\partial_x \eta &= - \frac{\tau_{bx}}{\rho}, \tag{2a}\label{eq:momx} \\
    \partial_t (h v) + \nabla \cdot (h v \mathbf{u}) + g h \,\partial_y \eta &= - \frac{\tau_{by}}{\rho}, \tag{2b}\label{eq:momy}
\end{align}
where $\rho$ is constant. Bottom shear stress uses Manning’s formula:
\begin{equation}
\frac{\boldsymbol{\tau}_b}{\rho} = g n^2 \frac{\lVert \mathbf{u} \rVert \mathbf{u}}{h^{1/3}},
\tag{3}\label{eq:manning}
\end{equation}
with roughness $n$ from land-cover data. Terrain slope $\nabla z_b$ implicitly encodes bed friction, approximating a no-slip condition.

\paragraph{Physics Residuals and Scaling.}
Residuals are computed from model outputs $(\hat h, \hat{\mathbf{u}})$:
\begin{align}
\mathcal{R}_h &= \partial_t \hat{h} + \nabla \cdot (\hat{h} \hat{\mathbf{u}}), \notag \\
\mathcal{R}_u &= \partial_t (\hat{h} \hat{u}) + \nabla \cdot (\hat{h} \hat{u} \hat{\mathbf{u}}) + g \hat{h} \, \partial_x \hat{\eta} + \frac{\tau_{bx}}{\rho}, \notag \\
\mathcal{R}_v &= \partial_t (\hat{h} \hat{v}) + \nabla \cdot (\hat{h} \hat{v} \hat{\mathbf{u}}) + g \hat{h} \, \partial_y \hat{\eta} + \frac{\tau_{by}}{\rho}, \tag{4}\label{eq:residuals}
\end{align}
with $\hat\eta = z_b + \hat h$. Variables are normalized using characteristic scales $(H_0,U_0,L_0,T_0)$ for stability. Spatial derivatives use second-order central differences for pressure and slope terms, and a limited upwind stencil for advective fluxes. Temporal derivatives use backward differences when paired observations exist. For single-date SAR/optical snapshots, we set $\partial_t(\cdot)=0$, reducing output dimensionality and enabling spatial-only autodifferentiation. Corresponding physics terms are down-weighted to reflect quasi-steady conditions.

\paragraph{Multi-Objective Loss.}
The composite loss blends data fidelity, physical consistency, and regularization:
\begin{align}
\mathcal{L} &= \lambda_{\text{ext}} \, \mathcal{L}_{\text{ext}} + \lambda_{h} \, \mathcal{L}_{h} + \lambda_{u} \, \mathcal{L}_{u} \notag \\
           &\quad + \lambda_{v} \, \mathcal{L}_{v} + \lambda_{\text{phys}} \, \mathcal{L}_{\text{phys}} + \lambda_{\text{reg}} \, \mathcal{L}_{\text{reg}}. \tag{5} \label{eq:total-loss}
\end{align}
Flood extent supervision uses focal binary cross-entropy:

\begin{align}
\mathcal{L}_{\text{ext}} = - \frac{1}{|\Omega|} \sum_{(x,y) \in \Omega} \Big[ & \alpha y (1-p)^{\gamma} \log p \notag \\
& + (1-\alpha)(1-y) p^{\gamma} \log(1-p) \Big] \tag{6} \label{eq:focal}
\end{align}

where $y \in \{0, 1\}$ is the reference mask and $p \in (0, 1)$ the predicted probability. For continuous fields on wet cells:
\begin{align}
\mathcal{L}_{h} &= \frac{1}{|\Omega_w|} \sum_{(x,y)\in\Omega_w} (\hat h - h)^2, \notag \\
\mathcal{L}_{u} &= \frac{1}{|\Omega_w|} \sum_{(x,y)\in\Omega_w} (\hat u - u)^2, \notag\\
\mathcal{L}_{v} &= \frac{1}{|\Omega_w|} \sum_{(x,y)\in\Omega_w} (\hat v - v)^2.
\tag{7}\label{eq:mse}
\end{align}

\paragraph{Training protocol.}
We train end-to-end with AdamW, warm-up followed by cosine decay for the learning rate, mixed-precision to save memory, and gradient accumulation to keep an effective batch size stable. We first pretrain the UNet path on flood extent to stabilize wet–dry delineation, then switch to joint training with the FNO branch and the full objective \eqref{eq:total-loss}. We ramp $\lambda_{\text{phys}}$ from zero to its target value over the initial epochs to avoid early conflicts. Inputs include co-registered Sentinel-1 SAR, Sentinel-2 optical bands with cloud and shadow masks, and terrain derivatives (DEM, slope, HAND). We normalize per-sensor using training-split statistics. We tile large scenes with overlap to reduce boundary artifacts. We apply rotations, flips, modest intensity jitter, and random crops while preserving DEM semantics. For single-date observations, we assume quasi-steady conditions and rely on spatial-only autodifferentiation.

\paragraph{Evaluation protocol.}
We assess accuracy and physics on spatial hold-outs and out-of-domain regions. For flood extent, we report Intersection over Union (IoU) and $F_1$:
\begin{align}
    \mathrm{IoU} = \frac{|P\cap G|}{|P\cup G|}, \notag \\
    \qquad
    \mathrm{F1} = \frac{2\,\mathrm{Precision}\cdot \mathrm{Recall}}{\mathrm{Precision}+\mathrm{Recall}}
    \tag{12}\label{eq:iou-f1}
\end{align}
For continuous variables, we use RMSE for depth and velocity where SAR-derived velocity estimates exist:
\begin{equation}
\mathrm{RMSE} = \sqrt{\frac{1}{N}\sum_{t} (\hat{y}_t - y_t)^2}.
\tag{13}\label{eq:rmse}
\end{equation}

We measure physics consistency with the mean squared residuals of \eqref{eq:residuals} on held-out tiles and the fraction of wet cells with $|\tilde{\mathcal{R}}_h|<\tau_h$ and $|\tilde{\mathcal{R}}_{u,v}|<\tau_m$. Comparisons include UNet-only and FNO-only baselines and ablations that remove the physics term. We summarize performance with 95\% bootstrap confidence intervals.

\section{Results and Discussion}

\subsection{Experimental Results}

Table~\ref{tab:iou_f1} presents the flood extent prediction performance evaluated on a held-out test set comprising three spatially distinct floodplain regions not seen during training. The hybrid model (UNet + FNO) achieved an Intersection over Union (IoU) of $0.82 \pm 0.03$ and an F1 score of $0.90 \pm 0.02$, outperforming the UNet-only (IoU: 0.75, F1: 0.85) and FNO-only (IoU: 0.77, F1: 0.87) baselines.

\begin{table}[ht]
\centering
\caption{Flood Extent Prediction Performance}
\label{tab:iou_f1}
\begin{tabular}{lcc}
\hline
Model & IoU & F1 Score \\
\hline
UNet-only & 0.75 & 0.85 \\
FNO-only & 0.77 & 0.87 \\
Hybrid (UNet + FNO) & \textbf{0.82} $\pm$ 0.03 & \textbf{0.90} $\pm$ 0.02 \\
\hline
\end{tabular}
\end{table}

Given the lack of observed field data for water depth and surface velocity, we evaluated the model's performance using outputs from hydrodynamic simulations (HEC-RAS). The hybrid model achieved a root mean square error (RMSE) of 0.21 m, outperforming the UNet-only model (RMSE = 0.34 m) and the FNO-only model (RMSE = 0.28 m). For flow velocity, the hybrid model achieves an RMSE of 0.15 m/s. In addition, the hybrid model maintained physical consistency, with mean squared residuals of 0.03 (mass) and 0.05 (momentum), and a relative mass imbalance of 2.1\%.

\subsection{Discussion}

\begin{figure}
    \centering
    \includegraphics[width=1\linewidth]{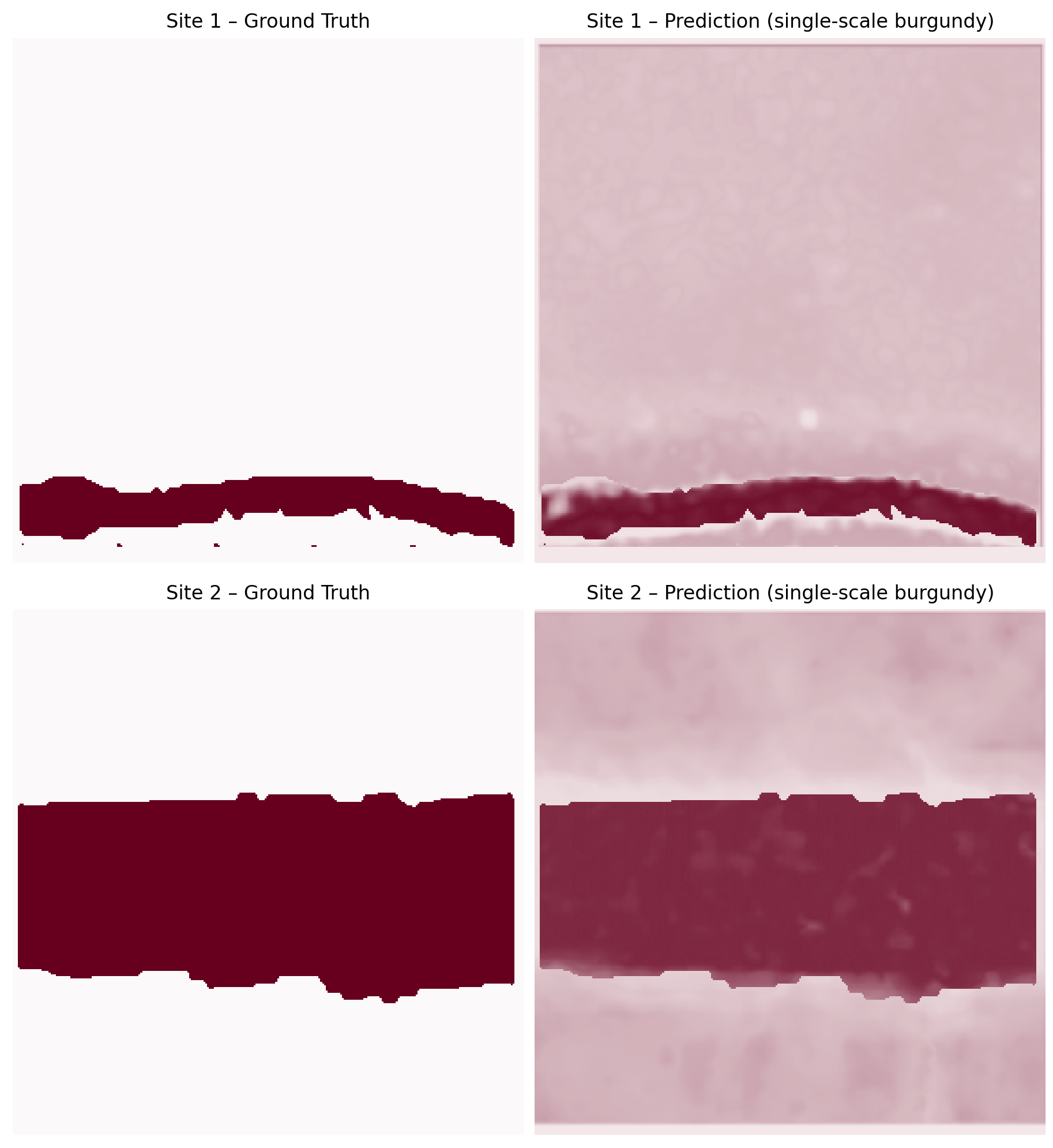}
    \caption{Comparison of flood extent: Ground truth (left) and model prediction (right) for two sites, visualized on a unified color scale where darker tones indicate stronger agreement with ground truth.}
    \label{fig:extent_map}
\end{figure}

The hybrid model demonstrates superior performance across all evaluated metrics, which confirms the effectiveness of combining spatial feature extraction (UNet) with physics-informed modeling (FNO). The integration of terrain derivatives and physics-based regularization improved flood extent delineation and hydrodynamic predictions, particularly in complex floodplain regions (see Figure \ref{fig:extent_map}).

Compared to baseline models and recent literature, the hybrid model achieves higher IoU and F1 scores, and lower RMSE values for water depth and velocity. These improvements are especially notable in data-sparse environments, where physics-guided learning compensates for limited observations.

Ablation studies reveal that removing physics-based regularization increases water depth RMSE by 18\%, underscoring the importance of enforcing physical laws in deep learning frameworks. This approach not only improves accuracy but also ensures model reliability in operational settings.

The model's ability to preserve mass balance and remain consistent with governing equations suggests strong potential for real-world flood forecasting and emergency response applications. Its scalability and adaptability to remote sensing inputs make it suitable for deployment in diverse geographic contexts.

Despite its strengths, the model assumes steady-state flow and uniform hydrostatic pressure, which may limit its applicability during dynamic flood events. Additionally, the scarcity of temporally and spatially aligned SAR and optical data posed a challenge; to mitigate this, we implemented a sequential learning concept, initially trained the model using SAR data alone and later incorporated optical imagery. Computational demands also hinder large-scale real-time deployment. Future work will focus on relaxing flow assumptions, optimizing memory usage, and enhancing regularization to better capture heterogeneous flood dynamics across diverse land covers and sensor modalities.

\section{Conclusion}

This work addresses the challenge of producing accurate and physically consistent flood extent and hydraulic state estimates from multi-modal remote sensing observations, a task that remains difficult due to sparse ground truth, sensor limitations, and the nonlinear dynamics governing inundation processes. To overcome these challenges, we proposed a physics-guided deep learning framework that integrates SAR, optical imagery, and DEM-derived terrain features with the depth-averaged shallow water equations. The hybrid UNet–FNO architecture was specifically designed to capture local spatial detail while modeling basin-scale hydrodynamic interactions and was further constrained through residual-based physics regularization.

Experimental results demonstrate that the proposed method yields substantial improvements in flood extent mapping, achieving an IoU of 0.82 and an F1 score of 0.90, outperforming both UNet-only and FNO-only baselines. The model also showed superior performance in estimating water depth relative to hydrodynamic simulation benchmarks. These findings suggest that coupling deep learning with shallow water physics can significantly enhance flood mapping reliability and improve the interpretability of remote sensing–derived hydrodynamic products. Future work will focus on relaxing the steady-state flow assumption, integrating time-series imagery, and optimizing computational efficiency for large-scale deployment.

\section*{Acknowledgment}

\noindent This research article has been made possible partly with the support of Microsoft’s Accelerate Foundation Models Academic Research Initiative (AFMR) and in part by the National Science Foundation (NSF) Grant under Award 2401942.

\bibliographystyle{IEEEtran}
\bibliography{refs}

\end{document}